\documentclass[aps,prl,twocolumn,showpacs]{revtex4}
\usepackage{graphics,graphicx}

\usepackage{color}
\usepackage{wrapfig}
\usepackage{enumerate}
\usepackage{amssymb,amsmath}
\usepackage{bm}
\usepackage[normalem]{ulem}

\begin{document}
\title{Magnetic phases in the S=1 Shastry-Sutherland model with uniaxial anisotropy}

\author{Lei~Su, Keola~Wierschem and Pinaki~Sengupta}
\affiliation{School of Physical and Mathematical Sciences, Nanyang Technological University, 21 Nanyang Link, Singapore 637371}

\date{\today}
%02.70.Ss Quantum Monte Carlo
%75.30.Kz Magnetic phase transition

\begin{abstract}
We explore the field induced magnetic phases of an $S=1$ $XXZ$  model with 
single-ion anisotropy and large Ising-like anisotropy on a Shastry Sutherland 
lattice over a wide range of Hamiltonian parameters and applied magnetic field. 
The multitude of ground state phases are characterized in detail in terms of their
thermodynamic properties and the underlying classical (Ising limit) spin arrangements 
for the plateau phases are identified by calculating the static structure factors.
The enlarged local Hilbert space of the $S=1$ spins results in several new ground state
phases that are not realized for $S=1/2$ spins. These include the quantum paramagnetic
state that is ubiquitous to $S=1$ spins with single ion anisotropy, two different 
spin supersolid phases (with distinct longitudinal ordering) and a magnetization
plateau that arises as a direct descendant of the 1/3 plateau due to
quantum fluctuations that are not possible for $S=1/2$ spins.
We predict the same mechanism
will lead to plateaus at smaller fractions of 1/3 for higher spins. The full momentum
dependence of the longitudinal and transverse components of the static structure 
factor is calculated in the spin supersolid phase to demonstrate the simultaneous 
existence of diagonal and off-diagonal long-range order as well as the different
longitudinal orderings.
\end{abstract}
\maketitle
\newpage
\section{INTRODUCTION}

The Shastry-Sutherland model (SSM) provides a paradigm to investigate the
emergence of novel quantum phases from the interplay between strong 
interactions, enhanced quantum fluctuations due to reduced dimensionality, 
geometric frustration and external magnetic fields~\cite{Shastry1981}.
 The model exhibits a a rich variety of ground
state phases with varying Hamiltonian parameters. 
The discovery 
of SrCu$_2$(BO$_3$)$_2$ and the observation of magnetization plateaus in
an external field garnered widespread interest and led to extensive investigation of the properties of the material and
the underlying SSM using both experimental and theoretical approaches~\cite{kageyama1999, miyahara1999, onizuka2000,kageyama2000, kodama2002, Miyahara2003}. Recently,
several additional spin systems have been shown to possess the same
underlying magnetic lattice, viz., the Shastry-Sutherland lattice. These include
a complete family of rare-earth tetraborides (RB$_4$, R = Tm, Er, Tb, Ho, Dy) \cite{Yoshii2006, Michimura2006, Siemensmeyer2008} and
the intermetallic Yb$_2$Pt$_2$Pb~\cite{Kim2008, Kim2013}. Preliminary investigations into the magnetic
properties of RB$_4$ compounds have revealed that the canonical SSM needs to 
be supplemented by additional longer range interactions to correctly account for
the behavior of these compounds~\cite{Suzuki2009,Suzuki2010}.

So far, the theoretical investigations of the SSM have almost exclusively focused on $S=1/2$ 
moments. Although the RB$_4$ compounds carry $S>1/2$ moments, the local Hilbert
space of the localized f-moments in many of these compounds is usually split by a strong single-ion
anisotropy into doublets and the low-energy physics is adequately described
by the lowest doublet in terms of an effective $S=1/2$ $XXZ$ model  with strong Ising-like exchange anisotropy.
The SSM with $S>1/2$ moments has remained  largely unexplored, although 
the importance of such extensions were realized in the original work of Shastry
and Sutherland. The enlarged Hilbert space of larger spins is expected to allow
for phases that are not found in the $S=1/2$ variant of the model. In this work, we
have addressed this by studying the $S=1$ Heisenberg model with uniaxial anisotropies 
on the Shastry-Sutherland lattice -- a straightforward generalization of the canonical 
Shastry-Sutherland model. In addition to
the aforementioned theoretical motivation, such a model is potentially
relevant to RB$_4$ compounds with an integer spin and  easy-plane single-ion anisotropy at moderate
temperatures -- large enough that thermal excitations to the first doublet are finite,
but small enough that contributions from the higher doublets are suppressed.

\section{Model}
\begin{figure}[t]
\includegraphics[width=2in,angle=0]{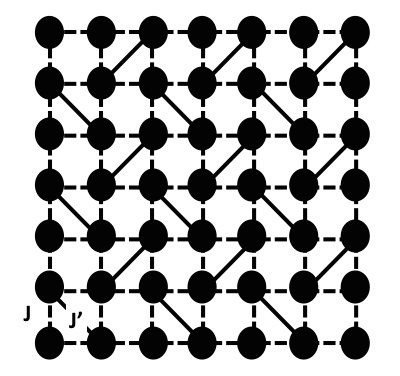}
\caption{
Shastry-Sutherland lattice. Interaction strengths between nearest neighbors and along alternating diagonals are $J$ and $J'$, respectively. 
}
\label{fig:SSL}
\end{figure}

The $S=1$ generalization of the Shastry-Sutherland model as studied here is 
 described by the  Hamiltonian
\begin{eqnarray}
{\cal H} &=& J\sum_{\langle i,j\rangle} \left[-\Delta\left(S_i^xS_j^x + S_i^yS_j^y\right) + S_i^zS_j^z\right] \nonumber \\
     & & + J'\sum_{\langle \langle i,j\rangle\rangle} \left[-\Delta\left(S_i^xS_j^x + S_i^yS_j^y\right) + S_i^zS_j^z\right] \nonumber \\
    & & + D\sum_i\left(S_i^z\right)^2 - B\sum_iS_i^z,
\label{eq:H}
\end{eqnarray}
where $\langle i,j\rangle$ and $ \langle \langle i,j\rangle\rangle $ refer to
summation over the nearest neighbors (nn) and along the diagonals of the 
Shastry-Sutherland lattice (SSL), 
respectively, and the corresponding interaction strengths are $J$ and $J'$ (Fig.~\ref{fig:SSL}). Henceforth, 
$J$ is set to unity and all the parameters are expressed in units of $J$. 
$\Delta$ is the measure of the exchange anisotropy - in this work, we consider
a strong and constant Ising-like exchange anisotropy, $\Delta=0.2$. $D>0$ measures the 
magnitude of easy-plane single-ion anisotropy,which is restricted to be easy-plane
in this study.  $B$ denotes an external 
longitudinal magnetic field. The exchange term is chosen to be ferromagnetic - this
eliminates frustration in the exchange part of the interaction and as a consequence,
the negative sign problem in quantum Monte Carlo (QMC) simulation is alleviated. Frustration is
retained in the Ising part of the interaction. In a real quantum magnet, both the
exchange and Ising part of the interaction have the same nature, so this may appear 
unphysical at first sight. But it was shown explicitly for the $S=1/2$ $XXZ$ model on the 
triangular lattice~\cite{Wang2009,Jiang2009} and the generalized $S=1/2$ Shastry-Sutherland model~\cite{Wierschem2013} (with 
additional longer range interactions) that in the Ising limit, the ferro- and anti-ferromagnetic
exchange interactions can be mapped on to each other. We expect similar arguments to
hold for the present model as well.
%Moreover, a ferromagnetic exchange appears naturally in the low energy effective model for a large-$S$ system with strong single-ion anisotropy.
%Finally, using a spin-boson mapping, the Hamiltonian (\ref{eq:H}) can be used to describe a system of interacting bosons on a SSL with large, but finite, on-site repulsion (Hubbard-$U$).

\begin{figure}[b]
\includegraphics[clip,trim=1.75cm 0cm 2.5cm 1cm,width=3in,angle=0]{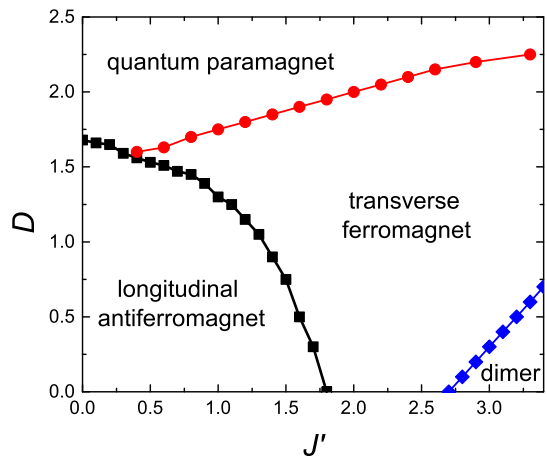}
\caption{
(Color online)
Ground state phase diagram of the $S=1$ Shastry-Sutherland model with easy-axis exchange anisotropy and easy-plane single-ion anisotropy. Transitions from longitudinal antiferromagnet to quantum paramagnet and to transverse ferromagnet are of first order while transitions from transverse ferromagnet to quantum paramagnet and to dimer state are continuous.
}
\label{fig:gs}
\end{figure}

In the limit of strong easy-axis single-ion anisotropy ($D<0$), 
a ferromagnetic exchange appears naturally in the low energy effective model for spin-$S$ systems with integer $S$~\cite{Wierschem2012}. 
This low energy effective model turns out to be the $S=1/2$ analogue of the model (\ref{eq:H}), 
and its ground state phases have been extensively explored~\cite{Meng2008}.
Later, the model was extended by including additional
longer range interactions in order to formulate an effective low energy model to capture
the magnetic properties of the rare-earth tetraboride family of quantum magnets~\cite{Suzuki2009,Suzuki2010,Wierschem2013}.

The $S=1$ model with isotropic antiferromagnetic interaction ($\Delta=-1$ in the present model) and
easy-axis single-ion anisotropy ($D<0$) was investigated in the framework of coupled chains of orthogonal dimers,
using exact diagonalization of small clusters and series expansion
methods much earlier in Ref.\cite{Koga2003}. In the limit of vanishing single ion anisotropy,
the ground state phase diagram is qualitatively similar to that of the canonical $S=1/2$
Shastry-Sutherland model: for small values of $J'$, the system has long range 
N{\' e}el (AFM) order whereas for large values of $J'$, the ground state is comprised
of singlets on the diagonal bonds.
However, the plaquette phase that is reported for
$S=1/2$ at intermediate values of $J'$, becomes unstable towards an alternative valence bond solid (VBS) ordering
for $S=1$ and it is not clear if the plaquette phase is stabilized for any finite range of 
parameters for the $S=1$ Shastry-Sutherland model.
A moderate value of $D$ was found 
to suppress both VBS phases completely.

For the case of ferromagnetic exchange considered in the present work, the singlet 
phase at large $J'$ is replaced by the triplet dimer phase - the ground state is comprised
of  triplets on the diagonal bonds. At small values of $J'$, the ground
state has long range Ising like AFM ordering - the SU(2) symmetry of the N{\' e}el state is
broken by the Ising-like anisotropy of the exchange term ($|\Delta| < 1$). A superfluid
ground state with transverse $XY$-ordering is stabilized at intermediate values of $J'$.

\section{Method and observables}

We have used the stochastic series expansion (SSE) quantum Monte Carlo (QMC) 
method with directed loop updates~\cite{Syljuasen2002} to simulate the Hamiltonian
on finite size lattices having dimensions $L\times L$, with $L = 4,\ldots 32$, in units
of the bare lattice spacing. Ground state behavior for finite lattices is accessed by
using sufficiently low temperatures - an inverse temperature of $\beta = 8\times L$
was found to be sufficient for the range of parameters studied. To characterize the
different ground state phases, we have studied the transverse ($S^{xy}(\bm q)$) and 
longitudinal ($S^{zz}(\bm q)$) components of the static structure factor,  
\begin{eqnarray} \label{eq:ssf}
S^{xy}(\bm q) &=& \frac{1}{N}\sum_{i,j}
	\langle S^x_i S^x_j + S^y_i S^y_j \rangle 
	e^{i{\bm q}\cdot({\bm r}_i-{\bm r}_j)}, \nonumber \\
S^{zz}(\bm q) &=& \frac{1}{N}\sum_{i,j}
	\langle S^z_i S^z_j \rangle 
	e^{i{\bm q}\cdot({\bm r}_i-{\bm r}_j)}.
\end{eqnarray}
These measure the degree of off-diagonal and diagonal ordering, respectively. Thus, in analogy to supersolids of lattice bosons~\cite{Boninsegni2012}, we can define spin supersolids \cite{Sengupta2007a,Sengupta2007b} by the simultaneous presence of transverse and longitudinal order.
A useful observable that detects the presence or absence of an excitation gap in any
phase is the spin stiffness. A finite stiffness accompanies a gapless state whereas 
a vanishing stiffness identifies a gapped state.  The spin stiffness $\rho_s$ is 
defined as the response to a twist in the boundary conditions~\cite{Sandvik1997}. 
In simulations that sample multiple winding number sectors, as in the present 
implementation of the directed loop algorithm, $\rho_s$ is simply related to 
the winding number of the world lines. For square simulation cells in two 
dimensions, $\rho_s=\left(w_x^2+w_y^2\right)/2\beta$, where $w_x$ and $w_y$ 
are the winding numbers in the $x$ and $y$ directions~\cite{Pollock1987}.
Finally, the uniform magnetization per site $m=\sum_j \langle S^z_j \rangle/N$ 
completes the list of observables. In order to probe the ground state properties 
of the different phases in the thermodynamic limit, we compute the various observables 
for different system sizes at multiple temperatures and extrapolate the results to the 
$ T \rightarrow 0 $ and $ L \rightarrow \infty$ limits.

\section{Results}

In this section we discuss the multitude of ground state phases that are stabilized 
for varying ranges of parameters of the model. In accordance with the argument 
provided earlier, we restrict our investigation to  $\Delta=0.2$ -- a representative 
value for a strong Ising-like exchange anisotropy.  We shall present the results in two parts -- 
first we discuss the ground
state phases over different ranges of the single-ion anisotropy ($D$) and the ratio
of diagonal to axial nn spin interaction ($J'$) in the absence of an external field. 
Subsequently,  the effects of a longitudinal field and emergent phases are described.   

\subsection{(i) Ground state phases at $B=0$}

The ground state phase diagram in the $J' - D$ parameter space, obtained 
from QMC simulations, is shown in Fig.~\ref{fig:gs}. A value $\Delta \neq -1$ explicitly 
breaks the SU(2) symmetry of the model. For small values of $J'$
and $D$, the system exhibits long range longitudinal staggered AFM order -- adiabatically connected 
to the ground state of the $S=1$ $XXZ$ model with no single-ion anisotropy on a square lattice ($J'=0,D=0$).
The interactions along the diagonals ($J'$) and the single-ion anisotropy ($D$) 
reduce the magnetic order, but do not suppress it completely at small values. The phase is 
identified by finite, non-zero values of the longitudinal component of the static
structure factor at the AFM ordering vector ${\bf Q}=(\pi,\pi)$. The transverse component 
is suppressed reflecting the absence of SU(2) symmetry. 
%In the limit of $J'=0$, the Hamiltonian ${\cal H}$ reduces to the $XXZ$ model on a square 
%lattice where ferromagnetic and anti-ferromagnetic exchange interactions are related
%to one another via a sublattice rotation. 
The two sublattices of the underlying square lattice (with the diagonals removed)
are populated by spins primarily in $S_i^z=+1$ and $S_i^z=-1$ states 
(up to quantum fluctuations). There is a gap to 
lowest spin excitations resulting in a vanishing spin stiffness.
%The single ion anisotropy opposes the AFM ordering and with increasing $D$, there is a discontinuous transition to a quantum paramagnetic (QP) phase with each spin predominantly in the $S_i^z=0$ state. This is distinct from the usual paramagnetic state with random spin orientation and is characterized by a finite value of the $zz$-component of the nematic tensor, ${\cal Q}^{zz}=\langle (S_i^z)^2-{2\over 3}\rangle$, that is induced by the single-ion anisotropy term.
Turning on the diagonal interaction, $J'$, opens up a new phases in the ground state phase diagram. 
As $J'$ is increased, keeping $D$ fixed (and small), the frustrated interaction reduces the 
magnetic order and eventually destroys it at a critical value via a first order quantum phase 
transition. The suppression of the longitudinal order is accompanied by the onset of
transverse ($XY$) ordering with a non-zero value of the transverse component of the 
static structure factor. The ferromagnetic exchange develops long range order and the ground 
state is marked by broken U(1) symmetry. This phase
is characterized by ferromagnetic ordering (driven by the exchange interaction) in the
transverse plane - the spin analog of the bosonic superfluid (SF) state.  Analogous to the emergence of Goldstone modes in bosonic superfluids, 
the gap to lowest excitations 
vanishes and is reflected in a finite stiffness. Upon increasing the strength of the diagonal 
interaction further at small $D$, 
the ground state undergoes a continuous quantum phase transition to a gapped
phase comprised of dimers along the diagonal bonds. This is the analog of the spin singlet
phase in the canonical Shastry-Sutherland model. The dimers in the present
model are not singlets, but $S=1$ triplets as a result of the ferromagnetic exchange 
interaction. However, it should be noted that unlike the singlet phase of the canonical 
model, a direct product of dimers does not constitute an exact eigenstate of the Hamiltonian, 
but is an approximation valid to leading order in $1/J'$\cite{Meng2008}. 

Next we consider the effect of varying the single-ion anisotropy on the nature of the ground states.
For small $J' (\lesssim 0.4)$, with increasing $D$, there is a discontinuous 
transition from the AFM state to a quantum paramagnetic (QP) phase at a critical value. 
In the QP phase, each spin is predominantly in the $S_i^z=0$ state, due to the energy cost associated with the $S_i^z=\pm 1$ states for $D>0$. This phase is distinct from the usual paramagnetic state with random spin orientation and is characterized by a finite value of the $zz$-component of the nematic tensor, ${\cal Q}^{zz}=\langle (S_i^z)^2-{2\over 3}\rangle$, that is induced by the single-ion anisotropy term~\cite{Zhang2013}. Long range order is  suppressed and there is a gap to lowest magnetic excitations.
%In the QP phase, the energy cost associated with $S_i^z=\pm 1$ states due to the large single-ion anisotropy results in the individual spins being primarily in the $S_i^z=0$ state. Long range order is  suppressed and there is a gap to lowest magnetic excitations.
For $J'\gtrsim 0.4$, the gapless SF phase is stabilized  
 over a finite range of $D$ intervening the AFM and QP phases.  
While the AFM-SF transition remains discontinuous, the SF-QP transition is a continuous one. 
Finally in the limit of strong interaction along the diagonal bonds ($J' \gtrsim 2.6$),
the dimer ground state at low $D$ gives way first to the SF phase and eventually 
to the QP phase via two continuous transitions. For intermediate
values of $J'$ $(1.8 \lesssim J' \lesssim 2.6)$, the SF phase extends all 
the way to $D=0$. We notice that, at $D=0$, the phase diagram resembles that 
of the $S =1/2$ model~\cite{Meng2008}. 

\begin{figure}[b]
\includegraphics[clip,trim=0cm 0cm 0cm 1cm,width=3.5in,angle=0]{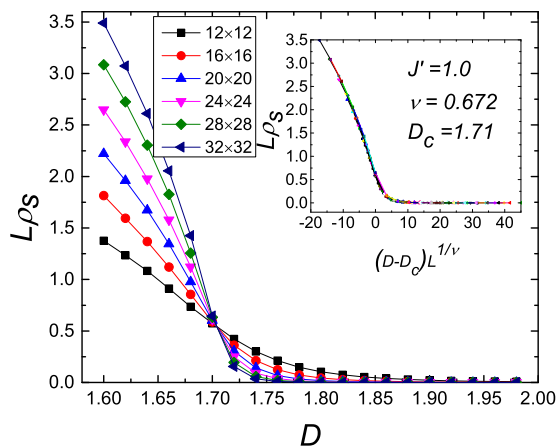}
\caption{
(Color online)
Finite size scaling for the spin stiffness $\rho_s$ at $J' =1.0$. The critical 
value $D_c$ is 1.71(1) and $\nu =0.672$.
}
\label{fig:scaling1}
\end{figure}

\begin{figure}[t]
\includegraphics[width=3.5in,angle=0]{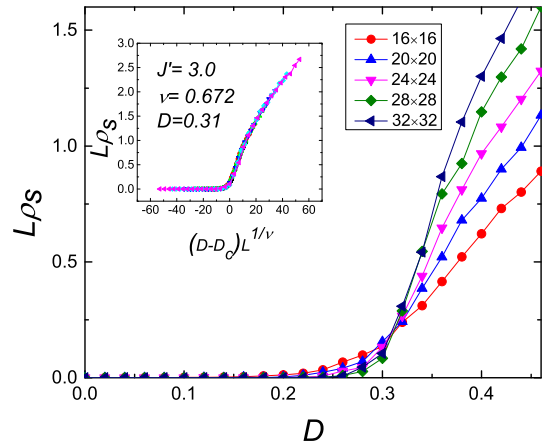}
\caption{
(Color online)
Finite size scaling for the spin stiffness $\rho_s$ at $J' =3.0$. The critical 
value $D_c$ is 0.31(1) and $\nu =0.672$.
}
\label{fig:scaling2}
\end{figure}

The transition between the Ising-AFM phase and the quantum paramagnetic phase 
-- both of which have a gap to lowest excitations -- is a discontinuous one. 
On the other hand, the dimer-SF and QP-SF transitions are driven by the breaking
of U(1) symmetry in the SF phase. Hence we expect these transitions to 
be continuous and belong to the O(2) universality class. This is confirmed
by the scaling of the observables at the associated phase boundaries. For example,
it is known from hyper-scaling theory that at a continuous phase transition, the spin 
stiffness for finite size lattices scale as 
\[
\rho_s(D,L)=L^{2-d-z}f\left({D-D_c\over D_c}L^{1/\nu},\beta/L^z\right),
\] 
where $z$ is the dynamical exponent which is unity for the O(2) universality class.
As $T\rightarrow 0$, $\rho_s(D,L)/L^{2-d-z}$ is a universal function of
${D-D_c\over D_c}L^{1/\nu}$ and the data for different system sizes collapse to a 
single curve. {\em At} the critical point ($D=D_c$), the ground state stiffness ($T=0$) scales as
\[
\rho_s(D_c,L)=L^{2-d-z}.
\]
that is, $\rho_s(D_c,L)/L^{2-d-z}$ is independent of system size and the curves for different
$L$ intersect at a point, providing an accurate estimate of the critical $D_c$.
Explicitly, the results are shown in Fig.~\ref{fig:scaling1} and Fig.~\ref{fig:scaling2}.

\subsection{(ii) Ground state phases at $B > 0$}

\begin{figure}[b]
\includegraphics[clip,trim=2.5cm 0.8cm 3.5cm 1.4cm,width=2.92in,angle=0]{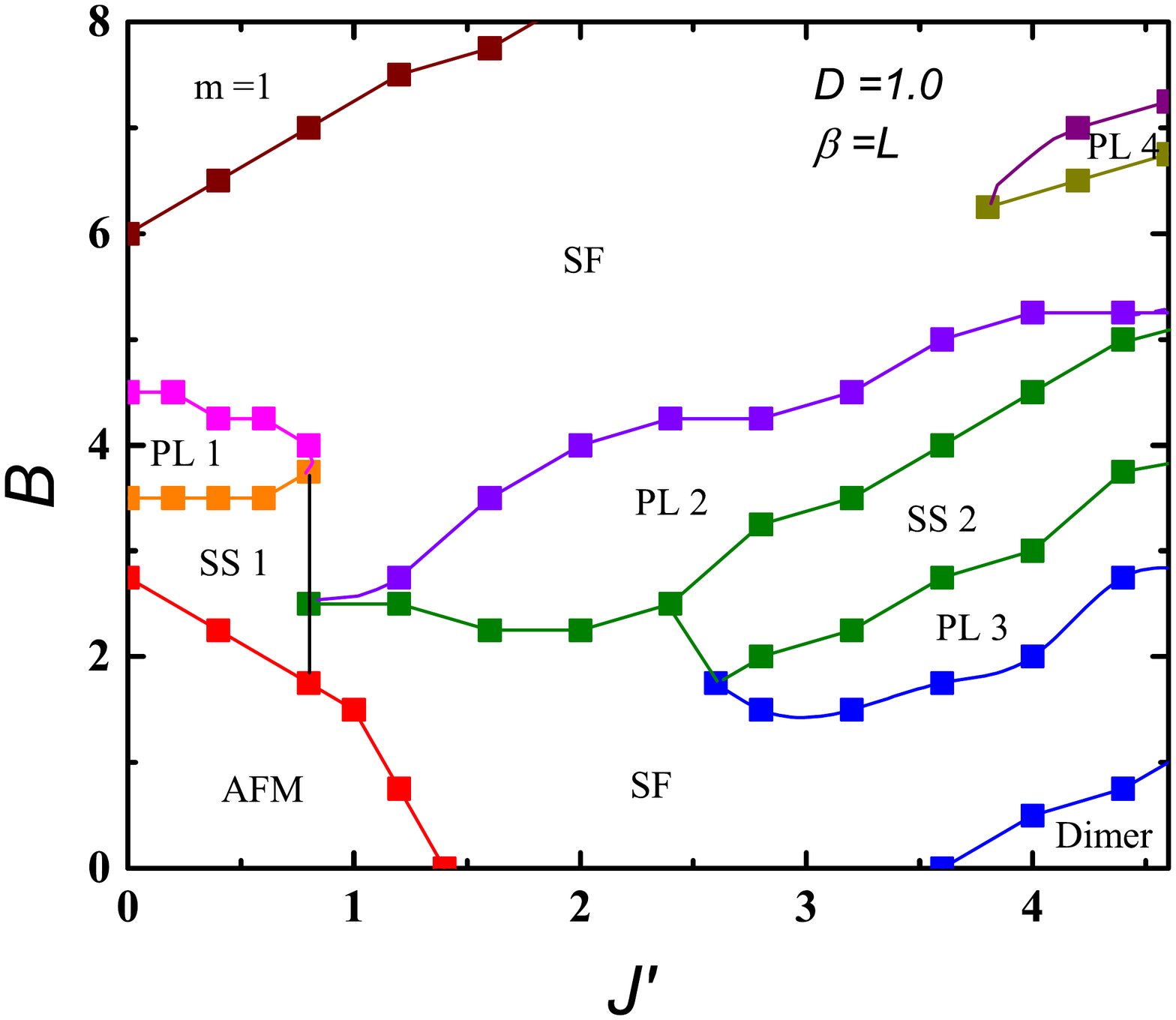}
\includegraphics[clip,trim=1.75cm 0cm 3.5cm 1cm,width=3in,angle=0]{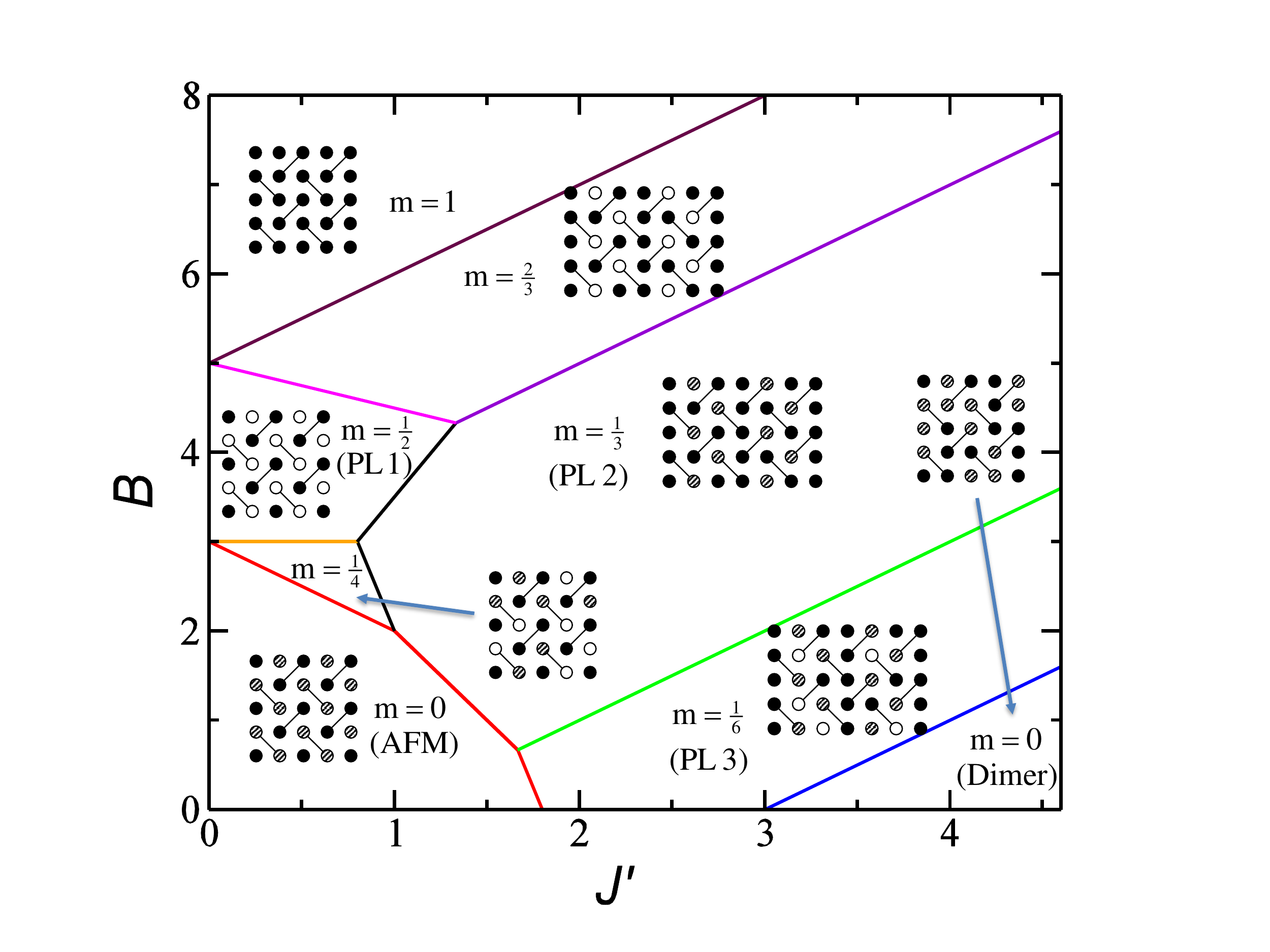}
\caption{ (Color online)
Upper Panel: Full phase diagram of the model at $\Delta=0.2$, $D=1.0$ and $\beta =L$. The application of an external longitudinal magnetic field, together with geometric frustration, results in the emergence
of a wide range of novel ground state phases, including magnetization plateaus (PL's) and supersolids (SS's).
Lower Panel: For comparison, the phase diagram in the Ising limit ($\Delta=0$) is also shown.
}
\label{fig:phd}
\end{figure}

The application of an external longitudinal magnetic field results in the emergence
of a wide range of novel ground state phases. The upper panel of Fig.~\ref{fig:phd} displays the full phase diagram, whose boundaries are obtained by comparing the energies of neighboring phases. In addition to the
phases observed at $B=0$, we see 
multiple magnetization plateaus (PL) with vanishing magnetic susceptibility, and 
spin supersolid (SS) phases where long range transverse magnetic order coexists with 
longitudinal magnetic order\cite{Liu1973}.
%(Keola: I think Liu and Fisher is more appropriate than Kim and Chan here)
%\cite{Kim2004}.
For small $J'$, the sequence of field-induced
phases is qualitatively similar to those obtained for the $XXZ$ model on a square 
lattice, viz. longitudinal AFM order with ${\mathbf Q}=(\pi,\pi)$ ordering - the
two sublattices are populated by spins in (predominantly) $S_i^z=+1$ and 
$S_i^z=-1$ states (modulo quantum fluctuations)\cite{Sengupta2007a}. The AFM 
state remains the ground state at small fields till the gap is closed at a 
critical field strength, beyond which a finite fraction of the spins are flipped 
from $S_i^z=-1$ to $S_i^z=0$ state, whereas the $S_i^z=1$ sublattice remains unaltered.
The ground state acquires additional  
long range transverse ordering via a continuous transition while still retaining the 
long range longitudinal AFM ordering. In other words, the ground state is a spin 
supersolid (SS1 in Fig.~\ref{fig:phd}). With increasing field, there is a transition 
to a longitudinally ordered
state with uniform magnetization $m=1/2$ (PL1). The ordering corresponds to 
replacing {\em all} the $S_i^z=-1$ spins on the down-sublattice  of the AFM state with 
$S_i^z=0$. The transverse order is completely suppressed and a gap opens up
in the excitation spectrum. The magnetization remains constant over a finite
range of field reflecting the spin gap -- this manifests itself as a plateau in the
$m-B$ curve. At still higher fields, the longitudinal order eventually melts 
and is replaced by an emergent transverse order. Finally, at very large fields,
there is a transition to the fully polarized state.

At larger $J'$, more interesting phases emerge due to stronger frustration. 
As the diagonal interaction increases to $J' \gtrsim 0.8$, 
%where the strength of diagonal bonds is equal to that of nn bonds, 
the SS1 and PL1 phases disappear. A new plateau emerges at $m=1/3$ (PL2). 
For $J'>1.5$, the AFM phase also disappears and the sequence of field-induced 
phases consists of SF-PL2-SF-FP. Upon further increasing  the diagonal
coupling to $J' \gtrsim 2.6$, two new phases appear -- a secondary plateau 
at $m=1/6$ (PL3) and a second kind of supersolid (SS2). Finally in the 
limit of large $J' (\gtrsim 3.6)$ the zero-field ground state enters the 
dimer phase -- long range magnetic order in the transverse plane is lost. 
A $m=1/2$ plateau (PL4) appears at high fields,
but the spin structure is markedly different from that in PL1.

%Ising-limit phase diagram
In order to better understand the nature of the plateau states, we also compare the energies of possible spin configurations in the Ising limit ($\Delta=0$) of each plateau and plot the phase diagram along with the local spin structures in Fig.~\ref{fig:phd}. In addition to the plateaus PL1, PL2 and PL3, as well as the zero-field AFM and Dimer phases, in the Ising limit there also exists a 1/4 plateau and a 2/3 plateau. The 1/4 plateau is obtained from the 1/2 plateau PL1 by changing the spin states on the $J'$ bonds from $(0 0)$ to $(\downarrow 0)$. However, as quantum fluctuations are added to the model, the 1/4 plateau becomes unstable towards the formation of the supersolid phase SS1, and by $\Delta=0.2$ it has vanished completely. Similarly, the 2/3 plateau is unstable towards delocalization of the $S_i=0$ spins into a superfluid state. This superfluid state significantly reduces the extent of many of the remaining plateaus. Between the 1/6 and 1/3 plateaus, the second supersolid phase SS2 emerges. PL4, which does not occur in the Ising limit for $D=1$, emerges at $\Delta=0.2$ as a gapped quantum disordered state.

As an illustration of the evolution of the  ground state across the multiple 
phase boundaries, Fig.~\ref {fig:four} summarizes the variation of several 
observables characterizing the different ground state phases as the external 
field is varied for representative values of the single-ion anisotropy 
($D = 1.0$) and diagonal interaction ($J' =4.0$). The longitudinal 
magnetization exhibits extended plateaus at 
$m = 1/6, 1/3$ and $1/2$ where the magnetization remains constant 
over a finite, non-zero field range. As a consequence of its short extent, 
the plateau at $m=1/2$ appears rounded. In the other phases (SF and SS), 
the magnetization grows monotonically. 
%At the plateaus, the ground state has long range diagonal order as indicated by the finite value of the longitudinal component of the static structure factor $S^{zz}({\bf Q})$, where ${\mathbf Q} =(2\pi/3,0)$ for the plateaus at $m=1/3$ and 1/6 (Fig. \ref{fig:four}(b)).
For the plateaus at $m=1/3$ and 1/6, the ground state has long range diagonal order 
as indicated by the finite value of the longitudinal component of the static structure 
factor $S^{zz}({\bf Q})$, where ${\mathbf Q} =(2\pi/3,\pi)$ or $ (\pi,2\pi/3)$ in consideration of symmetry breaking (Fig. \ref{fig:four}(b)). 
In contrast, the nature of the plateau at $m=1/2$ (PL4) is quite distinct from the other plateaus 
(including the other $m=1/2$ plateau, PL1) -- there is no long range magnetic order 
as evidenced by the vanishing structure factors (transverse and longitudinal). 
This state is adiabatically connected to a direct product state of 
${1\over \sqrt{2}}(|\uparrow0\rangle + |0\uparrow\rangle)$ 
along the diagonal $J'$ bonds. However, like the 
AFM state (and contrary to the dimer state in the $S=1/2$ SSM), the state is not a true eigenstate 
and the macroscopic degeneracy of the direct product state is lifted by quantum fluctuations. 
The presence of a finite gap to lowest excitations 
(and the absence of off-diagonal long range order) is manifested by vanishing 
values of the spin stiffness, $\rho_s$, and the transverse  component of the static 
structure factor, $S^{xy}({\mathbf Q})$ for all plateaus (Fig. \ref{fig:four}(c)(d)). 
Conversely, in the range of field strengths {$0.6 \lesssim B \lesssim 2.0$ and 
$5.2 \lesssim B \lesssim 6.4$, the ground state has long range 
off-diagonal order in the transverse plane (XY-AFM) marked by finite  $\rho_s$ 
and $S^{xy}({\mathbf Q})$. Over the intervening 
range of fields in between the $m=1/6$ and 1/3 plateaus -- $3.7 \lesssim B \lesssim 4.3$ -- the 
ground state is characterized by simultaneous diagonal and off-diagonal long 
range order. In other words, the ground state is a spin supersolid.

\begin{figure}[b]
\includegraphics[width=3.3in,angle=0]{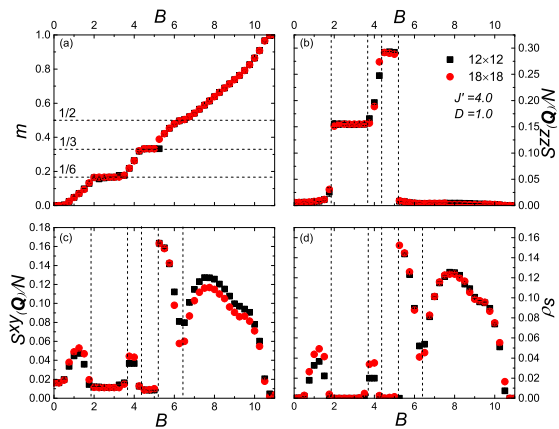}
\caption{
(Color online)
Evolution of the observables with an increasing external field at $J'=4.0$ and $D=1.0$. Lattice sizes are $L=12$ and $L=18$ respectively and reciprocal temperature $\beta = L$. (a) Magnetization. The $1/6$ and $1/3$ plateaus are prominent; the $1/2$ plateau is forming. (b) Longitudinal component of the structure factor $S^{zz}$ at ${\mathbf Q} =(2\pi/3,\pi)$ or $(\pi, 2\pi/3)$. (c) Transverse component of the structure factor $S^{xy}$ at ${\mathbf Q}$. (d) Spin stiffness as a function of magnetic field. Sandwiched between the $1/6$ and $1/3$ plateaus ($3.7 \lesssim B \lesssim 4.3$) is a supersolid phase which is characterized by finite $S^{zz}$, $S^{xy}$ as well as nonvanishing $\rho_s$. Error bars are smaller than the symbol size. 
}
\label{fig:four}
\end{figure}

\begin{figure}[t]
\includegraphics[width=3in,angle=0]{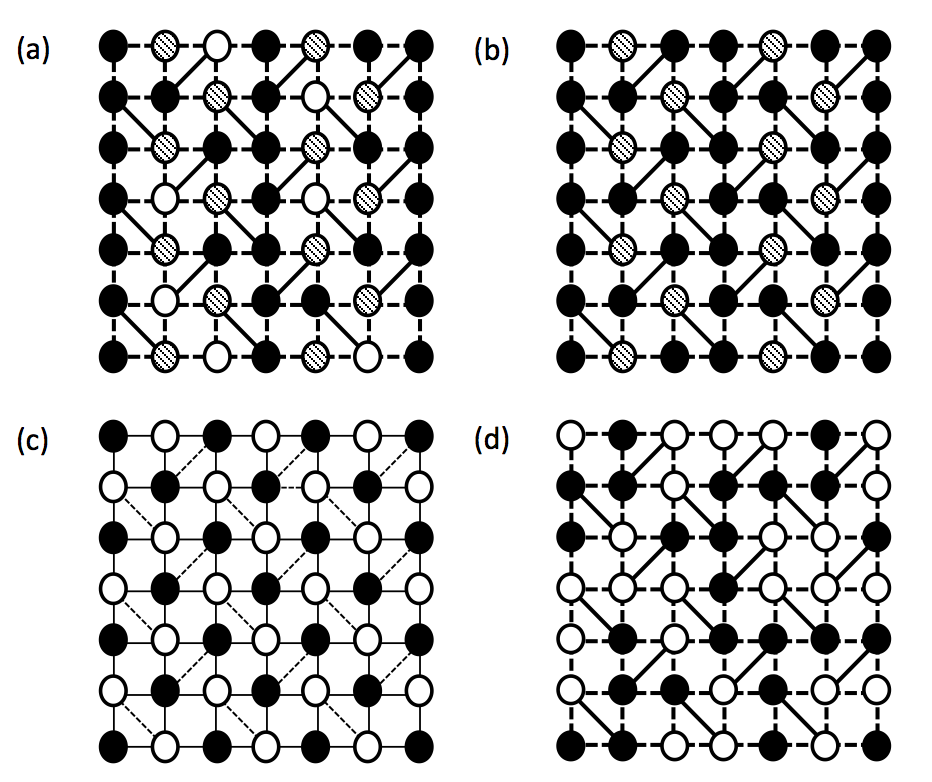}
\caption{
Schematic structures of the magnetization plateaus. (a) 1/6 plateau (PL3 in Fig.~\ref{fig:phd}); (b) 1/3 plateau (PL2); (c) 1/2 plateau (PL1) with small $J'$; (d)1/2 plateau (PL4) with large $J'$. The 1/6 plateau and the 1/3 plateau are characterized stripe structures. The 1/6 plateau is
obtained from the 1/3 plateau simply by replacing the $S^z=2$ $(\uparrow \uparrow)$ stripes with $S^z=1$ $(\uparrow 0)$ stripes. The 1/2 plateau in (c) has long range AFM order while the one in (d) is formed by dimers in a superposition of $S^z=1$ $(\uparrow 0)$ states. Black circles represent $\uparrow$; shaded $\downarrow$; white 0.
}
\label{fig:plat}
\end{figure}

\section{discussion}

The emergence of the $m=1/6$ plateau and the accompanying spin-supersolid phase are the 
most interesting results of the present study. These phases arise as a direct 
consequence of the extended local Hilbert space of $S=1$ spins and are absent 
in the $S=1/2$ model. To illustrate this, we show in Fig.~\ref{fig:plat} the spin arrangements 
in the $m=1/3$ 
and 1/6 plateaus in the Ising limit. Like their $S=1/2$ counterpart, the plateaus 
consist of identical dimers of strongly coupled spins along the diagonal bonds 
(with $S^z = 0, 1,$ or $2$) --  driven by large $J'$ -- arranged in stripes parallel to one 
of the principal axes (spontaneously breaking the $C_4$ symmetry). The magnetization 
at each plateau determines the modulation of the dimers. The 1/3 plateau 
consists of  stripes of $S^z=+2$ dimers  $(\uparrow \uparrow)$  and $S^z=0$ dimers 
 $(\uparrow \downarrow)$ arranged in a regular pattern with periodicity 3 - each period contains two 
$(\uparrow \downarrow)$ stripes and one $(\uparrow \uparrow)$ stripe. A peak in the longitudinal 
structure factor at ${\mathbf Q}$  confirms this picture. The 1/6 plateau is 
obtained from the 1/3 plateau simply by replacing the $S^z=2$ $(\uparrow \uparrow)$ stripes 
with $S^z=1$ $(\uparrow 0)$ stripes, without disturbing the periodicity of the diagonal order -- 
confirmed, once again, by a finite $S^{zz}(2\pi/3,\pi)$ or $S^{zz}(\pi, 2\pi/3)$. The appearance of such 
``offspring plateau'' is a direct consequence of the larger Hilbert space of $S=1$ spins 
and is not realized for $S=1/2$ spins.
For small values of $\Delta$, we anticipate the 1/2 plateau 
to be similarly accompanied by a 1/4 plateau, as occurs in the Ising limit. Additionally,
it would be interesting to study the competition between the SS2 phase and the putative $m=1/4$ plateau,
formed as an offspring of the 1/2 plateau PL4, in the intervening field range between the $m=1/3$ and 1/6 plateaus.
Unfortunately, our simulations were not suitable to probe the very strong interactions (large $D$ and $J'$) and
anisotropies (small $\Delta$) needed, thus precluding a direct investigation.

%We anticipate the 1/2 plateau to be similarly accompanied by a 1/4 plateau. 
%It would be interesting to study the competition between the SS2 phase and any putative $m=1/4$ plateau in the intervening field range between the $m=1/3$ and 1/6 plateaus.
%Unfortunately, our simulations were not suitable to probe the very strong interactions needed, thus precluding a direct investigation.

The magnetization process going from the 1/6 plateau to the 1/3 plateau with increasing 
field involves replacing the $(\uparrow 0)$ dimers by $(\uparrow \uparrow)$  dimers
{\em continuously}, without altering the $(\uparrow \downarrow)$ dimers. At intermediate values of the net
magnetization, $1/6 <  m < 1/3$, a finite number of $(\uparrow 0)$ dimers of the
1/6 plateau are replaced by $(\uparrow \uparrow)$. The stripes with $(\uparrow \downarrow)$
dimers remain intact, keeping the diagonal order unchanged. The ground state gains energy by 
(higher order) delocalization of the extra $(\uparrow \uparrow)$ dimers on the sublattice of the 
$(\uparrow 0)$ dimers, thus acquiring off-diagonal long range order. As a result, the ground
state possesses simultaneous diagonal and off-diagonal long range order. In other words, the 
ground state in this parameter range is a spin supersolid. The underlying mechanism of formation
of the spin supersolid phase is similar to that reported earlier for $S=1$ spins on a bipartite lattice~\cite{Sengupta2007a}
and bosonic supersolid phase for soft-core bosons with longer range interactions~\cite{Sengupta2005}.

In conclusion, we have demonstrated the emergence of several new phases in a Spin-1 Shastry-Sutherland model. These include the ``off-spring plateaus'' described above, as well as two new spin supersolid phases. These new phases appear as a direct result of the enlarged Hilbert space of the $S=1$ spins compared to their $S=1/2$ counterparts.

\begin{acknowledgments}
We thank Cristian Batista for useful discussions. L.S. acknowledges the support from the CN Yang Scholars Program at Nanyang Technological University.  This research used resources 
of the National Energy Research Scientific Computing Center, which is supported 
by the Office of Science of the U.S. Department of Energy under Contract No. DE-AC02-05CH11231. 
\end{acknowledgments}

\bibliographystyle{apsrev}
\bibliography{ref}

\end{document}